\documentclass[prb,twocolumn,showpacs,preprintnumbers,amsmath,amssymb]{revtex4}
\usepackage{graphicx}
\begin{document}

\title{Stability of antiphase line defects in nanometer-sized
boron-nitride cones}

\author{ S\'ergio Azevedo$^{1,2}$}
\author{ M\'ario S. C.  Mazzoni$^{1}$}
\author{ R. W. Nunes$^{1}$}
\author{H. Chacham$^{1}$}
\email{chacham@fisica.ufmg.br}
\affiliation{$^1$ Departamento de F\'{\i}sica, ICEX, Universidade Federal de Minas Gerais, CP 702, 30123-970, Belo Horizonte, MG, Brazil.}
\affiliation{$^2$ Departamento de F\'\i sica, Universidade Estadual de
Feira de Santana, Km 3 BR-116, 44031-460, Feira de Santana, BA, Brazil.}

\date{\today}

\begin{abstract}
We investigate the stability of boron nitride conical sheets of
nanometer size, using first-principles calculations. Our results
indicate that cones with an antiphase boundary (a line defect that
contains either B-B or N-N bonds) can be more stable than those
without one. We also find that doping the antiphase boundaries with
carbon can enhance their stability, leading also to the appearance of
localized states in the bandgap.  Among the structures we considered,
the one with the smallest formation energy is a cone with a
carbon-modified antiphase boundary that presents a spin splitting of
$\sim$0.5 eV at the Fermi level.
\end{abstract}

\pacs {71.20.Tx, 71.15.Mb, 71.24.+q}

\maketitle

\section{Introduction}
Boron nitride (BN) and carbon can be found in similar structures, from
the diamond and graphite bulk forms, to nanostructures such as
nanotubes,~\cite{iijima,chopra,zettl,loiseau} fullerenes
\cite{smalley,golberg} and conical sheets.~\cite{erik,bourgeois}
However, unlike carbon, BN structures can contain three types of
covalent bonds, namely B-N, N-N, and B-B.~\cite{zhang, dumitraca} This
leads to a qualitative difference between BN and carbon in the
formation of the graphite-derived curved surfaces that are found in
nanostructures such as fullerenes and nanotube caps.  In carbon,
energy minimization leads to structures made of 60$^\circ$
disclinations (conical sheets) \cite {foot1} with isolated pentagons
at the apex. Carbon fullerenes are built from such structures.  In BN,
fullerene-like structures are formed from 120$^\circ$ disclinations.
~\cite{golberg} These 120$^\circ$ BN disclinations are usually modeled
as having a four-membered ring at the apex,
\cite{golberg,jensen,simone1} but theoretical calculations indicate
that non-stoichiometric BN fullerenes, consisting of pentagon pairs
replacing the four-membered rings, may be energetically favored,
depending on the stoichiometric condition of
growth.~\cite{simone2,fowler}

A possible reason for the existence of 120$^\circ$ disclinations in BN
nanostructures, instead of the 60$^\circ$ ones that are found in
carbon, is the existence of antiphase boundaries (APB) in BN
disclinations with an odd multiple of
60$^\circ$.~\cite{azevedo,loiseau2} The common perception is that
these APB's, which necessarily contain a line defect of non-BN bonds,
should present an energy cost that surpasses the lower elastic-energy
cost of a disclination of smaller angle.  However, such a
``no-wrong-bond'' rule \cite{jensen} has been challenged by at least
two experiments.  In their work, Bourgeois {\it et al.}
\cite{bourgeois} have shown the existence of BN conical nanostructures
with a disclination angle of 300$^\circ$, demonstrating the existence
of APB's.  Another example in the literature is the observation of
disclinations in BN nanotube junctions, ~\cite {golberg2} in
experiments whose interpretation is also compatible with the existence
of an APB.

In the present work, we investigate the relative stability of
nanometer-sized BN cones with and without APB's, for N-rich and B-rich
environments.  Our first-principles results indicate that, for these
two limiting stoichiometric situations, structures with an APB can be
more stable than APB-free structures, consisting only of B-N bonds.
We also find structures involving full carbon incorporation at the APB
to be among the most stable, under both environments.

The paper is organized as follows: in Sec.~\ref{method} we discuss the
{\it ab initio} methodology for the computation of the total energies
and the definition of chemical potentials for the computation of the
zero-Kelvin grand potentials; in Sec.~\ref{apb-el} we discuss our
model for the formation energy as a sum of antiphase boundary and
elastic energies; in Sec.~\ref{abi} we discuss our first principles
results for the relative stability of the various cones and some
features of the electronic states of the most stable structures; and
finally, in Sec.~\ref{concl} we present some discussion of scaling to
larger structures than considered in this work and conclusions.

\section{Methodology}
\label{method}
\subsection{Total energy calculations}
Our calculations are based on the density functional theory
\cite{kohn} as implemented in the SIESTA program. \cite{siesta1} We
make use of norm-conserving Troullier-Martins pseudopotentials \cite
{troullier} in the Kleinman-Bylander factorized form,~\cite {bylander}
and a double-$\zeta$ basis set composed of numerical atomic orbitals
of finite range.  Polarization orbitals are included for nitrogen,
boron and carbon atoms, and we use the generalized gradient
approximation (GGA) \cite {gga} for the exchange-correlation
potential. All the geometries are fully relaxed, with residual forces
smaller than 0.1 eV/\AA.

\subsection{Zero-Kelvin grand potentials}
The comparison of BN clusters with different number of atoms is a
difficult problem. Here, we propose a zero-temperature thermodynamic
approach based on the prior determination of chemical potentials to
address this issue. The robustness of the approach will be established
by comparison with results obtained with different chemical
potentials, which represent different experimental conditions, and
with similar calculations in the
literature.~\cite{simone2,walter1,walter2,walter3,fazzio}

In order to address the energetics of stoichiometric and
nonstoichiometric cones, we must distinguish the different techniques
that can be employed in their synthesis. In each case, we introduce
suitable theoretical chemical potentials for nitrogen
($\mu_{N}^{\vphantom{B}}$) and boron ($\mu_{B}^{\vphantom{B}}$). For
the case in which the cones are obtained by laser ablation or
arc-discharge from hexagonal BN, we can have either a nitrogen- or a
boron-rich environment, depending on the specific atomic reservoir
employed.  In the N-rich environment, $\mu_{N}^{\vphantom{B}}$ is
obtained from nitrogen in the gas phase, while a metallic $\alpha$-B
phase is used as the reservoir for the B-rich environment.  In both
cases, $\mu_{N}^{\vphantom{B}}$ and $\mu_{B}^{\vphantom{B}}$ are
linked by the thermodynamic constraint
\begin{eqnarray}
\label{thermo}
\mu_{N}^{\vphantom{B}}+\mu_{B}^{\vphantom{B}}=\mu_{BN}^{layer}\;, 
\end{eqnarray}
where $\mu_{BN}^{layer}$ is the chemical potential per BN pair in the
infinite, planar BN sheet.  This constraint corresponds to situations
in which the nanostructures are formed in equilibrium with BN layers.
Since we are dealing with finite clusters, we use hydrogen atoms to
saturate the dangling bonds at the edges. To take these additional H-B
and H-N bonds into account, we introduce the respective chemical
potentials, $\mu_{HB}^{\vphantom{B}}$ and $\mu_{HN}^{\vphantom{B}}$,
and write the formation energy of the cones as
\begin{eqnarray}
\label{eform}
E_{form} = &&E_{tot} - n_{B}^{\vphantom{B}} \mu_{B}^{\vphantom{B}} -
n_{N}^{\vphantom{B}} \mu_{N}^{\vphantom{B}} - n_{HN}^{\vphantom{B}}
~\! \mu_{HN}^{\vphantom{B}} \nonumber\\
&&-n_{HB}^{\vphantom{B}}
~\! \mu_{HB}^{\vphantom{B}}\;,
\end{eqnarray}
where $E_{tot}$ is the calculated total energy of the cluster,
$n_{B}^{\vphantom{B}}$ and $n_{N}^{\vphantom{B}}$ are the number of B
and N atoms, and $n_{HN}^{\vphantom{B}}$ and $n_{HB}^{\vphantom{B}}$
are the number of H-B and H-N bonds, respectively.  The formation
energy, as defined in Eq.~\ref{eform}, represents the zero-Kelvin
grand potential which is the proper thermodynamical potential for the
comparison of the relative stability of structures with different
number of atoms.

A first constraint on the hydrogen chemical potentials is imposed by
using a finite planar sheet of boron nitride as reference, and
ascribing a null value to its formation energy.
This allows us to write its total energy as
\begin{eqnarray}
E_{T}^{sheet} = n_{BN}^{\vphantom{B}} ~\! \mu_{BN}^{layer} +
n_{HB}^{\vphantom{B}} ~\! \mu_{HB}^{\vphantom{B}} + n_{HN}^{\vphantom{B}}
~\! \mu_{HN}^{\vphantom{B}}\;,
\end{eqnarray}
where $n_{BN}^{\vphantom{B}}$ is the number of BN pairs.

The planar reference sheet is centered at a sixfold ring and has
$C_{3v}$ symmetry. Therefore, $n_{HB}^{\vphantom{B}} =
n_{HN}^{\vphantom{B}}$ and the preceding equation can be rewritten as
\begin{eqnarray} 
E_{T}^{sheet} = n_{BN}~\!\mu_{BN}^{layer} + n_{H}^{\vphantom{B}}
\mu_{H}^{\vphantom{B}}\;,
\end{eqnarray}
where we have defined $\mu_{H}^{\vphantom{B}} =
\left(\mu_{BH}^{\vphantom{B}} + \mu_{NH}^{\vphantom{B}}
\right)/2$. Using the total energy calculations for the infinite layer
and the finite sheet, we find $\mu_{H}^{\vphantom{B}} = -15.46~{\rm
eV}$.

From the definition of $\mu_H^{\vphantom{B}}$, we obtain
$\mu_{HB}^{\vphantom{B}}$ ($\mu_{HN}^{\vphantom{B}}$) once
$\mu_{HN}^{\vphantom{B}}$ ($\mu_{HB}^{\vphantom{B}}$) is known. The
values of one of these parameters can be determined by choosing a
convenient reservoir. In our calculations we use the ammonia molecule
(NH$_3$) as a reservoir, obtaining $\mu_{HN}^{\vphantom{B}} =
-16.16$~eV and $\mu_{HB}^{\vphantom{B}} = -14.76$~eV in the N-rich
environment, and $\mu_{HN}^{\vphantom{B}} = -15.23$~eV and
$\mu_{HB}^{\vphantom{B}} = -15.69$~eV in the B-rich environment.  The
consistency of this methodology was tested by determining
$\mu_{HB}^{\vphantom{B}}$ using the BH$_3$ molecule as a reservoir,
and then calculating $\mu_{HN}^{\vphantom{B}}$ from
$\mu_H^{\vphantom{B}}$. The calculated formation energies in this case
are within $0.01$~eV/atom of those from the NH$_3$ reservoir.  For the
structures containing carbon, the chemical potentials for the carbon
atom ($\mu_{C}^{\vphantom{C}}$) and the carbon-hydrogen bonds
($\mu_{CH}^{\vphantom{C}}$) are obtained from total energy
calculations for bulk graphene and for a finite carbon sheet saturated
by hydrogen atoms, respectively.

\section{Formation energy: antiphase-boundary and elastic contributions}
\label{apb-el}
The physical mechanism that defines whether structures having APB's
may be as stable as their APB-free counterparts is the competition
between the elastic energy associated with the cone formation and the
energetic cost of forming an APB.  Hence, we write the cone formation
energies, calculated with the above procedure, as the sum of two
contributions. The first is the elastic energy required to fold a
finite sheet into a cone, which we describe by a continuum model.  The
second represents the APB chemical energetic cost associated with the
non-BN bonds.

The basic assumption of the continuum model is that the elastic energy
stored in a differential element of area at the surface of the cone is
proportional to the square of the surface mean curvature in that
region. Within this model, the elastic energy of the cone is written
\begin{equation}
\label{Eel}
E_{el} = K \left(\frac{\cos^{2}\!\!{\alpha}}{\sin{\alpha}}\right)
\ln\left(\frac{R}{R_c}\right) + E_{tip}(R_c)\;,
\end{equation}
where $2\alpha$ is the apex angle, $R$ is the slant height,
$R_c$ is a cutoff length introduced to remove the singularity at the
tip (linear elasticity breaks down in the neighborhood of the tip),
and $E_{tip}(R_c)$ is the associated tip energy. The expression
\begin{equation}
\sin(\alpha) = 1 - \frac{\beta}{2\pi} = 1 - \frac{N}{6}
\end{equation}
relates the apex angle to the disclination angle $\beta = N\times
60^\circ$. 

The validity of the continuum model was first tested for the case of
carbon, for which there is no APB. We found that the first-principles
total energy calculations for the graphene sheet, and for the cones
with $60^\circ$ and $120^\circ$ disclinations are fitted remarkably
well by a straight line, when plotted as a function of
$\cos^{2}\!\!{\alpha}/\sin{\alpha}$, as shown in Fig.~1.
\begin{figure}[h]
\includegraphics[width=0.40\textwidth]{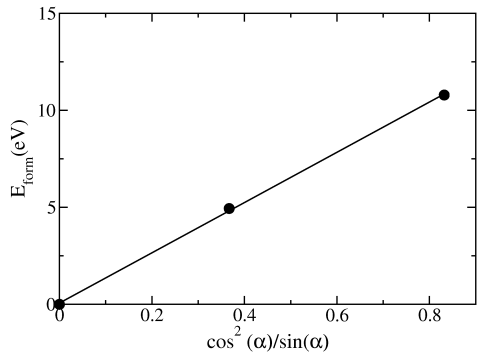}
\caption{\label{fig:ajuste} Formation energy as a function of 
$\cos^{2}\!\!{\alpha}/\sin{\alpha}$ for carbon cones with $60^\circ$
and $120^\circ$ disclinations and for the graphene sheet.} 
\end{figure}
This fact also shows that there is no tip contribution to the
formation energy, i.e. $E_{tip} =0$, at least for structures with
disclination angles up to 120$^\circ$. This can be explained by the
fact that these geometries have the general appearance of truncated
cones. Figure~1 also allows us to consider that $R_c$ is the same for
the cones with 60$^\circ$ and 120$^\circ$ disclinations.

Adding the APB energies, we write 
\begin{eqnarray}
E_{form}^{\vphantom{B}}= K
\left(\frac{\cos^{2}\!\!{\alpha}}{\sin{\alpha}}\right)
\ln\left(\frac{R}{R_c}\right) + E_{APB}^{\vphantom{B}}\;,
\end{eqnarray}
for the formation energy of BN cones.  The constant of proportionality
$K$ can be obtained simply by considering that for a $120^{\circ}$
disclination there is no APB, hence $E_{APB}^{\vphantom{B}} =
0$.~\cite{foot2} Having $K$ and the formation energies, we can use the
equation above to estimate the APB energies.

\section{First principles results}
\label{abi}
Figure 2 shows three of the most stable structures found in our
study. Note that each contains a 60$^\circ$ disclination, leading to
the formation of an APB.  We considered two kinds of APB's: one
contains a sequence of parallel (N-N or B-B) bonds, as shown in
Figs. 2(a) and 2(c), and the other a sequence of zig-zag (N-N or B-B)
bonds. These are denoted molecular APB and zig-zag APB,
respectively.~\cite{azevedo} For cones with 120$^{\circ}$
disclinations, we considered structures with a four-membered ring at
the apex (without non-BN bonds), as well as structures with two
pentagons at the apex, adjacent or not. Disclinations of 180$^{\circ}$
and 300$^{\circ}$ (each containing an APB) were also addressed. To
make contact with experimental results for the synthesis of BN cones
from carbon templates, we also considered the possibility of carbon
incorporation in BN cones, leading to the formation of carbon zig-zag
[shown in Fig. 2(e)] and molecular APB's. In these structures, every
(B or N) atom along the APB is replaced by a carbon atom.

\vspace{0.2cm}

\begin{figure}[h]
\includegraphics[width=0.50\textwidth,bb=80 180 500 710,clip]{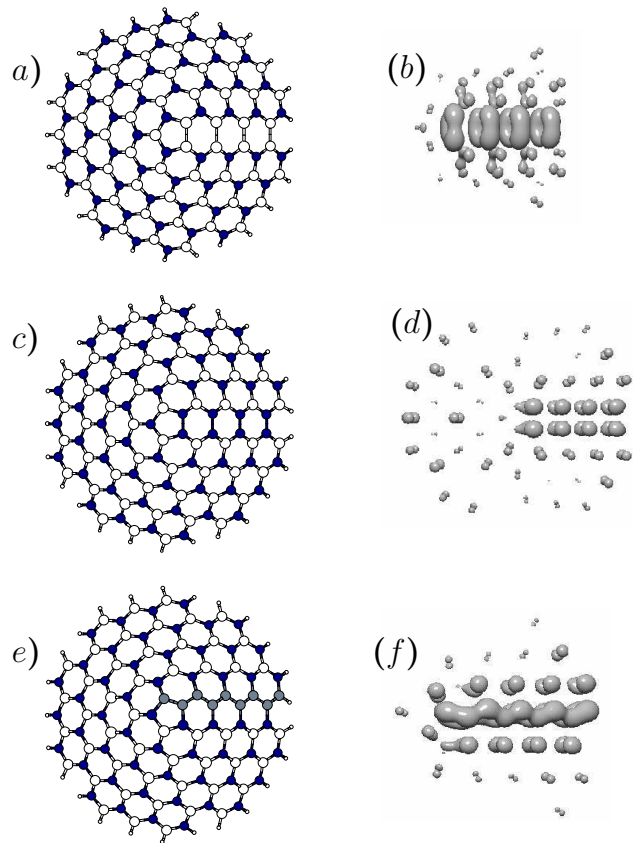}
\caption{\label{fig:cones} (Color online) Stable boron nitride
nanometer-sized conical structures with $60^\circ$ disclinations and
an APB (see text). Boron, carbon, and nitrogen atoms are represented
by white, gray, and black (blue) circles, respectively. Hydrogen atoms
saturate the dangling bonds at the edges. (a) and (b) show a molecular
APB of boron and the electron distribution of its empty gap states,
respectively. (c) and (d) show the same for a nitrogen molecular APB
and its filled gap states. (e) and (f) show the same for a zig-zag APB
of carbon and its filled gap states.}
\end{figure}

Our first principles results for the formation energy of cones with
various disclination angles, for B-rich and N-rich environments, are
shown in Table I.  For the 60$^\circ$-disclination cones, the third
and fourth columns of Table~I show the ratio $E_{APB}^{\vphantom{B}} /
E_{el}^{\vphantom{B}}$, where $E_{el}^{\vphantom{B}} = 2.96~{\rm eV}$,
from the continuum model.  Note that the highest APB energies are
associated with zig-zag configurations, more specifically a zig-zag
boron APB in the N-rich environment, and a zig-zag nitrogen APB in the
B-rich environment. Geometric aspects may explain these
results. Indeed, zig-zag APB's introduce larger strains in the
structures when compared to molecular APB's, since the parallel
arrangement of B-B and N-N bonds in the latter is more easily
accommodated by the BN matrix of the cone.  Table I also shows that
for some structures the energetic cost of creating an APB is
comparable to the elastic energy, i.e.  $E_{APB}^{\vphantom{B}} /
E_{el}^{\vphantom{B}} \sim 1$.  For the B-rich environment, these are
the molecular APB of boron [shown in Fig.~2(a)] and the zig-zag APB of
carbon (with C-B bonds), while for the N-rich environment, the ratio
is $\sim$1 for the molecular APB of nitrogen [shown in Fig.~2(c)] and
the zig-zag APB of carbon (with C-N bonds)[shown in Fig.~2(e)].  From
these observations, we expect that cones with these APB's will be the
best candidates to compete energetically with APB-free cones.
 
Our results for the formation energies, shown in the last two columns
of Table I, confirm this expectation.  As discussed in
Ref.~\onlinecite{simone2}, in a condition of chemical equilibrium, the
relative stability of the structures is determined by the quantity
$E_{form}/n_{poor}$, where $n_{poor}$ is the number of atoms of the
species for which the environment is poor [$n_B^{\vphantom{B}}$
($n_N^{\vphantom{B}}$) in the N-rich (B-rich) case]. The underlined
numbers in Table~I indicate the most stable structures. A general
trend that can be inferred from the numbers in table I is that the
existence of non B-N bonds does not necessarily increase the formation
energy. The stability of non-stoichiometric cones depends strongly on
the environment.  This conclusion is valid for isolated non B-N bonds
as well as for the different kinds of APB's.

\begin{table}
\caption{Formation energies (in eV) of stoichiometric and non-stoichiometric
cones, for N-rich and B-rich environments. The first column
indicates the disclination angle for each structure.  Second column
indicates either the type of APB or the topological defect at the
apex, in each case.  For the 120$^\circ$ and 180$^\circ$
disclinations, we considered different possibilities for the defect at
the apex (indicated by the number of squares and pentagons in each
case).  The third and fourth columns show the ratio between the the
antiphase-boundary ($E_{APB}$) and elastic energies ($E_{el}$) for the
60$^\circ$ disclinations.  The last two columns show the formation
energy per $n_{poor}$, as discussed in the text.  Underlined numbers
indicate the most stable structures for each environment.}
\label{tab1}
\begin{tabular}{ccccccccc}
\hline
Discl. & Cone &  
& ${E_{APB}^{\;\!}\over{E_{el}}}^{\;\!}$
& ${E_{APB}^{\;\!}\over{E_{el}}}^{\;\!}$
& ${E_{form}\over {n(B)}}$
& ${E_{form}\over{n(N)}}$\\
&&& (N-rich)    & (B-rich) &(N-rich)& (B-rich) \\ \hline
$120^\circ$ & 1 sq.&   &  &    & 0.15  & 0.15  \\
$120^\circ$   &2 pent. &  &  &  & 0.17 & 0.30\\
$120^\circ$   &2 ad. pent. &  &  &   &0.12& 0.14  \\
\hline
$60^\circ$   & Zig-N &  &  2.9 & 7.9 & 0.21 & 0.44 \\
$60^\circ$   & Zig-B &  &  7.7 & 2.7 & 0.43 & 0.20  \\
$60^\circ$   & Mol-B &  &  3.8 & 1.3 & 0.24 & \underline{0.12} \\
$60^\circ$   & Mol-N &  &  1.1 & 3.8 & 0.11 & 0.24  \\
$60^\circ$ &Zig-C (C-B) &  & 4.4 & 1.2 & 0.29 & 0.13  \\
$60^\circ$ &Zig-C (C-N) &  & 0.6 & 3.8 & \underline{0.09} & 0.26 \\
$60^\circ$ &Mol-C (C-B) &  & 7.1 & 2.1 & 0.43 & 0.18  \\
$60^\circ$ &Mol-C (C-N) &  & 1.4 & 6.0 & 0.14 & 0.37  \\
\hline
$180^\circ$   & 3 pents &  &  &   &0.48 & 0.85  \\
$180^\circ$   & 1 pent. + 1 sq. &  &  &   & 0.42 & 0.66  \\
\hline
$300^\circ$ & Mol-N &  &  &    &  1.44 & 1.61

\end{tabular}
\end{table}

Concentrating on the structures with 120$^\circ$ disclinations, we
note that, under N-rich conditions, the energy of a cone with two
non-adjacent pentagons (containing three N-N bonds) is comparable to
that of a cone with a four-membered ring at the apex.  The same is
true if the pentagons are adjacent. The latter is a stoichiometric
structure with only one non-BN bond, and its formation energy is
independent of the environment.

Concerning the APB's in geometries with 60$^\circ$ disclinations, we
remark again that under the two limiting stoichiometric conditions we
considered, 60$^\circ$ disclinations with $E_{APB}^{\vphantom{B}} /
E_{el}^{\vphantom{B}} \sim 1$ [the molecular APB of nitrogen (in the
N-rich case) and the molecular APB of boron (in the B-rich case)] can
be more stable than the APB-free structures with 120$^\circ$
disclinations. These are stable structures, despite being responsible
for the existence of gap states.~\cite{azevedo} (Gap states are absent
in the case of cones with a 120$^{\circ}$ disclination.)

The electronic charge densities associated with the gap states for the
molecular APB's of boron and nitrogen are shown in Figs.~1(b) and
1(d), respectively.  The nitrogen-APB states show a substantial degree
of delocalization, spreading over the BN matrix of the cone, while the
boron-APB states remain strongly localized on the B atoms at the
APB. In a previous study,~\cite{azevedo} we examined the electronic
levels of these two structures. In the nitrogen case, we found that
three fully occupied levels appear in the lower half of the gap, and
Fig.~1(d) shows that they have the same nitrogen lone-pair character
that dominates the states at the top of the valence band of the
cone. This leads to a strong mixing between the APB states and the
valence-band states in the bulk of the cone, this being the reason for
the delocalized character of the nitrogen-APB states.  In the case of
boron, the states shown in Fig.~1(b) are empty levels near the bottom
of the conduction band. They have a more metallic character, differing
from the states at the bottom of the conduction band. Mixing is absent
in this case, leading to the localized distribution seen in Fig.~1(b).

Table~I also shows that cones with zig-zag APB's have higher formation
energies, and the same occurs for structures with large disclination
angles. The result is expected in the case of the 180$^\circ$ and
300$^\circ$ disclinations, since besides the cost of forming an APB,
the elastic energy is also large. However, we note that these
metastable structures have already been found
experimentally.~\cite{golberg2,bourgeois}

Regarding carbon-doped cones, we find that incorporation of carbon at
the APB leads to highly stable structures. In this case, for both
kinds of APB's the most stable structures are those where the APB is
connected to the BN matrix by C-N (C-B) bonds in the N-rich (B-rich)
environment. Actually, the most stable of all structures in our study
is the C-N connected carbon zig-zag APB, in the N-rich environment,
shown in Fig. 1(e). The electronic charge density of the gap states
associated with this structure is shown in Fig.~1(f). The states are
predominantly localized in the antiphase boundary, strongly resembling
the states characteristic of the graphite $\pi$ band. This
wavefunction localization leads also to a considerable spin splitting
at the Fermi level. For the neutral charge state, the energy
difference between the highest-occupied molecular orbital (which has
majority spin) and the lowest-unoccupied one (which has minority spin)
is $\sim 0.5~{\rm eV}$. This spin splitting appears only for
structures with an odd number of electrons, a result that raises the
interesting question about what may happen in the limit of cones with
larger sizes.  Furthermore, we have also identified a tendency for
carbon atoms to segregate at the APB. In our calculations, a single
pair of carbons atoms is found to be more stable (by $\sim$0.07-0.15
eV/atom) on the APB than in the bulk of a BN cone. This may indicate a
tendency for the formation of these carbon-modified APB's, during the
synthesis of BN conical nanostructures by substitution on a carbon
template.

Boron nitride cones can also be obtained by a substitution reaction
\cite{golberg2} on a carbon template in which B$_2$O$_3$ and gaseous
N$_2$ are used as sources of boron and nitrogen, respectively. To
check the stability trends in this case, we consider that BN cones and
CO$_2$ are the by-products of this reaction.  The nitrogen and boron
chemical potentials are then obtained in terms of the total energies
of the species involved.~\cite {reaction} With this procedure, the
value we obtained for the chemical potential difference, $\mu_N -
\mu_B = -189.936$~eV, is very close to the value of -190.126~eV,
obtained for an N-rich environment under the constraint in
Eq.~\ref{thermo}.  Therefore, the stability trends shown in table~I
still hold in this case.  We also checked the robustness of our
results with respect to changes in the boundary conditions, which
could happen, for instance, if the cones were interacting with
different surfaces. This was done by imposing a constant shift of
$\pm$ 0.5~eV on all hydrogen chemical potentials. The ordering of the
quantities $E_{form}/n_{poor}$ in table~I remains unaltered, with
relative shifts of less than 0.02 eV.  Still to confirm the validity
of our methodology, we mention that similar approaches have been
employed in the literature in the determination of formation energies
of defects in cubic boron nitride,~\cite{walter1,walter2,walter3} in
boron nitride nanotubes \cite {fazzio} and in the investigation of
possible structures of boron nitride fullerenes.~\cite {simone2} In
the latter work, the stability trends obtained by using the solid
phases of boron and nitrogen together with the constraint in
Eq.~\ref{thermo} remained largely unchanged when the authors
considered the gas phase limit of atomic reservoirs.
 
\section{Remarks on size scaling and conclusion}
\label{concl}
Finally, we discuss the relevance of our results in face of the
expected scaling trends. As the size of a cone increases, the APB
energy scales linearly with the slant height $R$ (see Eq.~\ref{Eel}),
i.e. with the square root of the total number of atoms in the
structure, while the elastic energy scales sub-linearly with $R$, as
shown in Eq.~\ref{Eel}. Hence, structures without APB's become
energetically favored for sufficiently larger sizes.  However, cones
of the size we consider in this work may represent the seeds present
in the initial steps of the nucleation process that generates larger
cones or other nanostructures. Not only the energetics, but also
kinetics play an important role in this process, and to the extent
that we may expect low energy configurations to be favored for the
seeds, cones with certain types of APB's may be good candidates, and
kinetics may determine the formation of larger size structures
containing APB's.

To summarize, we used first-principles calculations to address the
question of how stable are antiphase boundaries in boron nitride
nanometer-sized conical nanostructures. Our results indicate that
60$^\circ$ disclinations with an APB can be more stable than APB-free
120$^\circ$ disclinations.  We show that the formation energies of the
stable APB's are comparable to the disclination elastic energy, and
that the dependence of the latter on the disclination angle is very
well described by a continuum model. Also, we observe that carbon
incorporation at the APB leads to highly stable structures, of which
the most stable presents a large spin splitting ($\sim 0.5~{\rm eV}$)
at the Fermi level.

\begin{acknowledgments}

We acknowledge support from the Brazilian agencies CNPq, FAPEMIG, and
Instituto do Mil\^enio em Nanoci\^encias-MCT.

\end{acknowledgments}


\begin{thebibliography}{100}
\bibitem{foot2} 
There is a very small difference in slant height between our
60$^\circ\!$- and 120$^\circ\!$-disclination structures, which leads
to a negligible change in the logarithmic term in Eq.~\ref{Eel}
between these two types of structures. This has been neglected in our
determination of $K$.
\bibitem{reaction}
The substitution reaction described allows us to write the nitrogen
and boron chemical potentials as $\mu_N={1\over2}E(N_2)$ and
$\mu_B={1\over2}[{E(B_2O_3) + {3\over2}E(C) - {3\over2}E(CO_2)]}$,
where E(N$_2$) and E(CO$_2$) are the total energies of the nitrogen
and carbon dioxide molecules, respectively, $E(B_2O_3)$ is the total
energy of a planar hexagonal $B_2O_3$ structure and E(C) is the energy
per atom in graphene.  All these quantities are obtained by means of
first-principles calcultions and the resulting values of $\mu_N$ and
$\mu_B$ are -270.151 eV and -80.214 eV, respectively.
\end{thebibliography}
\end{document}